\documentclass[a4paper]{jpconf}
\usepackage{amssymb, amsmath, iopams}
\begin{document}

\title{Nilpotence in Physics: the case of Tsallis entropy}

\author{A J Creaco$^1$, N Kalogeropoulos$^{2,}$\footnote[3]{Corresponding Author}}
\address{$^1$ Science Department, BMCC - The City University of New York, 199 Chambers St., New York, NY 10007, USA}
\address{$^2$ Pre-Medical Department, Weill Cornell Medical College in Qatar, Education City, PO Box 24144, Doha, Qatar}

\ead{\ $^1$ acreaco@bmcc.cuny.edu, \ $^{2,3}$ nik2011@qatar-med.cornell.edu}

\begin{abstract}
 In an attempt to understand the Tsallis entropy composition property, we construct an embedding of the reals into the set of 
 $3\times 3$ upper triangular matrices with real entries. We explore consequences of this embedding and of the  geometry of the 
 ambient $3\times 3$ Heisenberg group. This approach establishes the polynomial growth of the volume of phase space 
 of systems  described by the Tsallis entropy and provides a general framework for understanding Abe's formula in terms of 
 the Pansu derivative between Riemannian spaces.
\end{abstract}


\section{Introduction}
The Harvda/Charvat [1], Dar\'{o}czy [2], Tsallis [3] entropy (henceforth Tsallis entropy) is a single parameter family of  entropic functionals 
introduced in the Statistical Mechanics community in 1988 [3]. It has attracted considerable interest since that time, for several reasons. One 
is that it claims to describe phenomena having long range spatial and temporal correlations, something that the Boltzmann/Gibbs/Shannon (BGS)
entropy is, a priori, unsuitable to do [4] (and references therein). Examples are long-range interactions such as the gravitational interaction
applied, in particular, in astrophysical or general relativistic contexts and electrodynamics as applied in charged plasmas, for instance.
Another reason is that the Tsallis entropy appears to be suitable for describing some out of equilibrium phenomena. 
As it befits a concept claiming a foundational role in any part of Physics, as well as a wide-ranging set of applications, the 
Tsallis entropy has, inevitably, also attracted some criticism (see e.g. [5], [6]). Even though its scope of applicability, if any, is currently unclear, 
its contribution in providing a viable alternative to the BGS entropy which gives rise to a reasonable, and well-developed, 
thermodynamic formalism is undeniable.


\section{Additivity and extensivity}
Consider a set having a sample space $\Omega$. As examples of physical interest, $\Omega$ can be a countable set with a measure, such as the 
cardinality of subsets, and a metric, or it can be a Riemannian manifold or, more generally, a metric measure (mm) space. We will follow the notation of 
the  Riemannian case in the sequel, but the results can be straightforwardly extended to the discrete case or further generalized to mm spaces.  Let the 
statistical behavior of the aspects of the system under study be given by the probability function $p: \Omega \rightarrow [0,1]$. Assume that $p$ has no 
atoms and  let its Radon-Nikodym derivative  with respect to $dvol_\Omega$ be indicated by $\rho$. The Tsallis entropy of $p$ is defined [3],[4] as 
\begin{equation} 
     S_q [\rho ] = k_B \ \frac{1}{q-1} \left\{ 1 - \int_{\Omega} [\rho (x)]^q \ dvol_\Omega \right\}
\end{equation}
Here \ $k_B$ \ is Boltzmann's constant that will be set equal to one, for simplicity, in what follows. The parameter \ $q\in\mathbb{R}$, where $\mathbb{R}$
indicates the set of the real numbers,  is called nonextensive or entropic parameter. It is straightforward to check that   
\begin{equation}
     \lim_{q\rightarrow 1} S_q = S_{BGS}
\end{equation}
Consider two occurrences $\Omega_1, \Omega_2 \subset \Omega$  that are independent, in the conventional sense of the word, namely
$\rho_{\Omega_1 \ast \Omega_2} = \rho_{\Omega_1} \cdot \rho_{\Omega_2}$ where $\Omega_1 \ast \Omega_2$ stands for the combined system 
resulting from the interactions of $\Omega_1$ and $\Omega_2$. The Tsallis entropy is then 
\begin{equation}
    S_q [\rho_{\Omega_1 \ast \Omega_2} ] = S_q [\rho_{\Omega_1}] + S_q [\rho_{\Omega_2}] + (1-q) S_q [\rho_{\Omega_1}] S_q [\rho_{\Omega_2}]
\end{equation}
According to (3), the Tsallis entropy is not additive, in the conventional sense of the word. However this is not an impediment to formulating a 
valid thermodynamic formalism [4] (and references therein). Additivity is quite distinct from extensivity, although in the usual BGS entropy case
this distinction is somewhat blurred. The important property that any thermodynamic potential should have is extensivity, appropriately defined. This  
amounts to requiring that the thermodynamic potentials are proportional to the number of effective degrees of freedom of the system. Clearly, additive 
systems are extensive, but the converse is not necessarily true. Property (3) above is, arguably, the most distinct property of the Tsallis entropy when 
compared to its BGS counterpart. In order to restore the manifest additivity in (3), a generalized addition $\oplus_q$ was defined [7],[8] by
\begin{equation}
       x \oplus_q y = x + y + (1-q) x y, \ \ \ \forall \  x, y \in \mathbb{R}
\end{equation} 
This is a usual process in Physics, where new objects are defined to naturally reflect properties of some operations: consider the role of tensors, for 
instance, in General Relativity, the role of covariant derivatives in Gauge Theories and Gravity, appropriately defined superfields in supersymmetric and 
supergravitational models etc. Using (4) the generalized independence/additivity of the Tsallis entropy can be rewritten in the more ``covariant" form  
\begin{equation}
        S_q [\rho_{\Omega_1\ast \Omega_2}] = S_q [\rho_{\Omega_1}] \oplus_q S_q [\rho_{\Omega_2}]
\end{equation}


\section{BGS vs. Tsallis entropies: hyperbolic metrics}
In an attempt to compare the concept of independence as encoded through the BGS and the Tsallis entropies, we set up in [9] a coordinate system 
whose one axis was parametrized by $\mathbb{R}$ and whose other axis was parametrized by its Tsallis entropy induced deformation [10]. By using 
a semi-direct product construction, we found that the resulting plane has a metric of constant sectional curvature which is given by
\begin{equation}
     k = - [\log (2-q) ]^2
\end{equation}
If the Tsallis entropy composition property is a feature of phase space of a system, as opposed to being emergent (e.g  ultrametrics  in the case of 
spin glasses) then it induces a hyperbolic structure on phase space. Hence someone can claim that the Tsallis entropy is the ``hyperbolic analogue" of 
the ``Euclidean" BGS entropy [9]. Then the Cartan-Hadamard theorem guarantees that the Tsallis entropy is universal, if one also accepts the 
rest of the Shannon/Khitchine/Abe/Santos/Suyari axioms. The conclusion that we reached can be extended to the case of interacting systems via 
convex potentials, even when the systems are described tby different values of the entropic 
parameters, in which case, the resulting system is described by an induced $CAT(k)$ structure rather than by one induced by the hyperbolic plane.  
As a result of this formalism, we were able to establish in [11] that if a system's thermodynamic behavior is described by the Tsallis entropy and if the 
composition property relfects a property of its phase space, then the largest Lyapunov exponent of the underlying dynamical system should be zero. 
The converse is not necessarily true though. Hence all such systems described by the Tsallis entropy exhibit ``weak chaos". Moreover under the above 
assumptions we were able to show in [12] why one should use the escort distribution
\begin{equation} 
     \rho_{esc} = \frac{1}{N} \  \rho^q, \hspace{5mm} N = \int_{\Omega} \rho^q \ dvol_{\Omega}\end{equation}
rather than $\rho$ itself in maximizing and in deriving the moments of the quantities of interest stemming from the Tsallis entropy. 


\section{BGS vs Tsallis entropies: sub-Riemannian structures}
The goal of the construction of this section is the same as of the previous one: to compare the usual and the generalized additions in an attempt to 
elucidate the meaning and implications of the Tsallis entropy composition. Here, we construct a matrix the non-trivial elements of 
which give rise to the ordinary and generalized additions, upon matrix multiplication. Then we explore geometric consequences of this construction. 
To begin with, we perform a formal re-scaling  of the Tsallis entropy 
\begin{equation}
       \tilde{S}_q [\rho ] \equiv (1-q) S_q [\rho ]
\end{equation}
Then $\tilde{S}_q$ obeys a modified additivity which induces a generalized addition, mimicking the steps leading to (4), given by  
\begin{equation}
     x \tilde{\oplus}_q y = x + y + xy
\end{equation}
Next, we define the map  $\mathcal{S}: \mathbb{R} \rightarrow \mathfrak{U}^{3\times 3}$ defined by  
\begin{eqnarray}
    \mathcal{S} (x) \ = \  \left(      \begin{array}{ccc}
                                                          1 & x &  x \\
                                                           0 & 1 & x \\
                                                           0 & 0 & 1 
                                                      \end{array}                \right)
\end{eqnarray}
Here $\mathfrak{U}^{3\times 3}$ stands for the set of upper triangular matrices with real entries having unit elements in their diagonal.  
Upon ordinary matrix multiplication, we get
\begin{eqnarray}
  \mathcal{S}(x) \ \mathcal{S}(y) \ = & \left(
                                                                  \begin{array}{ccc}
                                                                     1 & x + y & x \tilde{\oplus}_q  y \\
                                                                     0 &    1    &  x + y         \\
                                                                     0 &    0    &     1
                                                                  \end{array} \right)      
\end{eqnarray}
This is a very desirable form since the elements in the secondary diagonal are additive, whereas the ``corner" term follows the generalized additivity 
(10). So, the embedding (11) allows someone to compare the ordinary and the generalized additions within the same structure.\\

A drawback of (11) is that it is not a homomorphism, nor is the image of $\mathbb{R}$ under $\mathcal{S}$  a subgroup of $\mathfrak{U}^{3\times 3}$. 
The latter is bothersome, on physical grounds, as we wish to have a structure that allows the composition of elements, such as a group or, more 
generally, a category. If we wish to continue working with groups, we are lead to consider not just with the image of $\mathbb{R}$ under $\mathcal{S}$, 
but the whole ambient space $\mathfrak{U}^{3\times 3}$. This group is easily checked to be 2-step nilpotent and its Lie algebra is easily seen to be 
generated by the matrices   
\begin{equation}
     X =   \left( 
            \begin{array}{ccc}
              0 & 1 & 0 \\
              0 & 0 & 0 \\
              0 & 0 & 0 
            \end{array} \right),   \hspace{7mm}
    Y =   \left(
           \begin{array}{ccc}
              0 & 0 & 0 \\
              0 & 0 & 1 \\
              0 & 0 & 0 
           \end{array} \right),   \hspace{7mm}
     Z =  \left(
            \begin{array}{ccc}
              0 & 0 & 1 \\
              0 & 0 & 0 \\
              0 & 0 & 0 
           \end{array} \right)         
\end{equation}
which obey $[X, Y] = Z$ with all other commutators being zero. This is exactly the 3-dimensional Heisenberg algebra. Hence someone can claim
that the Tsallis entropy can be viewed as the ``quantum analogue" of the BGS entropy, understanding however that this is just a formal analogy and not 
claiming any deeper physical significance to it, at least at this stage. The effective ``Planck's constant" in this description, is a function of $q$.\\

From a geometric viewpoint, the embedding (11) reduces the study of the Tsallis entropy composition (10) into the framework of the geometry of the 
Heisenberg groups, or more generally brings it in the far wider context of sub-Riemannian (Carnot-Carath\'{e}odory) geometry [13], [14]. 
This framework allows
us to draw some far-reaching conclusions. One of them is that if a system is described by the Tsallis entropy and if the composition property is not 
emerging, but it follows directly from a corresponding property of the phase space, then the phase space volume of the system grows in a  polynomial 
(power-law) fashion. This stems from a continuum analogue of a theorem due to Milnor/Wolf/Tits/Guivar'ch/Bass/Gromov stating that a finitely generated 
group has polynomial growth if and only if it is virtually nilpotent. The desired result  can be reached by either taking the ``continuum limit", or by applying 
Kanai's discretization results for manifolds with lower Ricci curvature bounds. The conclusion is an 
inverse, in a sense, of a result of Tsallis et al. [4] in the discrete case, and is applicable for both discrete and continuous systems. 
Another implication of this construction  is that we can see Abe's formula involving the Jackson derivative of a probability distribution as a special 
case of the Pansu derivative between sub-Riemannian spaces. Details and more concrete explanations  of the statements of this paragraph 
will be given in a forthcoming work.  


\ack{We wish to thank the Organizing Committee of IC-MSQUARE 2012 for inviting us to present our work in the mini-workshop 
``Gravity, Quantum, and Black Holes" and for establishing a pleasant and stimulating venue for the effective presentation and 
fruitful exchange of ideas.}


\end{document}